\documentclass{article}
\usepackage{bm}
\usepackage{amsmath}
\usepackage{graphicx}
\usepackage{amsfonts}
\usepackage{setspace}
\usepackage{lineno}
\usepackage{url}
\usepackage[version=4]{mhchem}

\title{Quantifying AI data center nitrogen oxide (\ce{NO_x}) emissions from space}
\author{
    Kevin D. Gauld\textsuperscript{1,*},
    Daniel J. Varon\textsuperscript{1,2,*},\\
    Nicholas Balasus\textsuperscript{3},
    and Daniel H. Cusworth\textsuperscript{3}\\[1ex]
    \footnotesize\textsuperscript{1}Massachusetts Institute of Technology, Department of Aeronautics and Astronautics\\[-0.75ex]
    \footnotesize\textsuperscript{2}Massachusetts Institute of Technology, Institute for Data, Systems, and Society\\[-0.75ex]
    \footnotesize\textsuperscript{3}Carbon Mapper\\[-0.75ex]
    \footnotesize\textsuperscript{*}Corresponding Authors: kgauld@mit.edu, dvaron@mit.edu
}
\date{}

\begin{document}

\maketitle

% \linenumbers

\begin{abstract}
AI data center power demand is spurring rapid deployment of on- and near-site natural gas turbines. The resulting nitrogen oxide (\ce{NO_x}) pollution is a growing concern but has not previously been quantified with atmospheric observations. Here we demonstrate space-based detection and quantification of \ce{NO_x} emissions from the SpaceXAI Colossus 2 power plant in Southaven, Mississippi. Using observations from the geostationary TEMPO satellite instrument, we detect a strong increase in local mean \ce{NO2} column concentrations after the plant began operations in late 2025. We then use TEMPO to estimate two-week-average \ce{NO_x} source rates from August 2025 to August 2026, calibrating against continuous emission monitoring system (CEMS) data from US power plants. TEMPO first detected \ce{NO_x} emissions in December 2025 at 530$\pm$220 kg h$^{-1}$. We find that emissions increased through August 2026, averaging 870$\pm$165 kg h$^{-1}$ after February 2026, roughly 19$\pm$4 times higher than expected from the facility’s March 2026 permit for 41 turbines operating under best available control technology (BACT) requirements ($\sim$47 kg h$^{-1}$). Emissions at the expected level would be undetectable by our TEMPO analysis.
\end{abstract}

\section*{Introduction}

The rapid expansion of AI data centers is increasingly reliant on natural gas turbines to meet electricity demand, raising concerns about emissions of toxic air pollutants including nitrogen oxides (\ce{NO_x}, consisting of \ce{NO} and \ce{NO2}), formaldehyde, and fine particulate matter \cite{NAACPvxAI2026}. These turbines can supplement power from the local grid and may be located on-site or at dedicated near-site facilities. In effect, some large AI data centers operate private gas power plants, with potentially limited emissions transparency and oversight. The resulting air pollution may significantly impact nearby communities \cite{NAACPvxAI2026} but has not yet been quantified observationally.

Here we demonstrate space-based detection and quantification of \ce{NO_x} emissions from the SpaceXAI Colossus 2 data center, the largest operational data center in the United States \cite{trinetti2026}, using observations from NASA’s TEMPO satellite instrument, which provides hourly coverage of the contiguous United States at 2$\times$4.5-km$^2$ resolution \cite{zoogman2017}. As of July 2026, Colossus 2 supported roughly 770,000 GPUs, with a total computing power demand of nearly 1 GW \cite{trinetti2026}. It is located in Memphis, Tennessee but draws power from a dedicated gas power plant roughly 1.5 km away in Southaven, Mississippi. SpaceXAI began deploying temporary trailer-mounted gas turbines to that location and operating them in late 2025, gradually increasing the number to 69 by August 2026 \cite{MCEQ2026}. In March 2026, a permit was issued for the construction of 41 permanent turbines equipped with selective catalytic reduction (SCR) systems to reduce \ce{NO_x} emissions \cite{MDEQ2026}, but delayed turbine buildout has led to continued reliance on the temporary units \cite{MCEQ2026}. As of late July 2026, only 14 of the 69 turbines were SCR-equipped \cite{MCEQ2026}, and the resulting emissions are of uncertain magnitude. We show here that TEMPO can detect and quantify the ramp-up of \ce{NO_x} emissions from the power plant over time, reaching levels far higher than expected from the permit. Our work demonstrates the capability of satellites to monitor large \ce{NO_x} emissions from data center power generation and assess regulatory compliance.

\section*{Data and Methods}

Satellite instruments in low-Earth orbit (LEO) have observed atmospheric \ce{NO2} column concentrations since the 1990s with spatiotemporal resolution approaching a few km and daily revisits. New instruments in geostationary orbit such as TEMPO now provide hourly km-scale \ce{NO2} monitoring across continental domains \cite{zoogman2017}. Here we use these observations to detect and quantify \ce{NO_x} emissions from the Colossus 2 power plant over time. We first compare oversampled TEMPO \ce{NO2} column concentrations near the power plant in the months before and after operations began, revealing a strong increase in local \ce{NO2} levels. We then average the hourly TEMPO observations over time with wind rotation to generate images of two-week-average \ce{NO2} plumes from the facility. From these plume images, we infer biweekly \ce{NO_x} emissions using a cross-sectional flux (CSF) method calibrated against continuous emission monitoring systems (CEMS, \cite{CAMPD}) data from seven US power plants. Details of the retrieval and emission quantification methods are provided in the Extended Methods.

\section*{Results and Discussion}

\begin{figure}%[tbhp]
\centering
\includegraphics[width=0.8\linewidth]{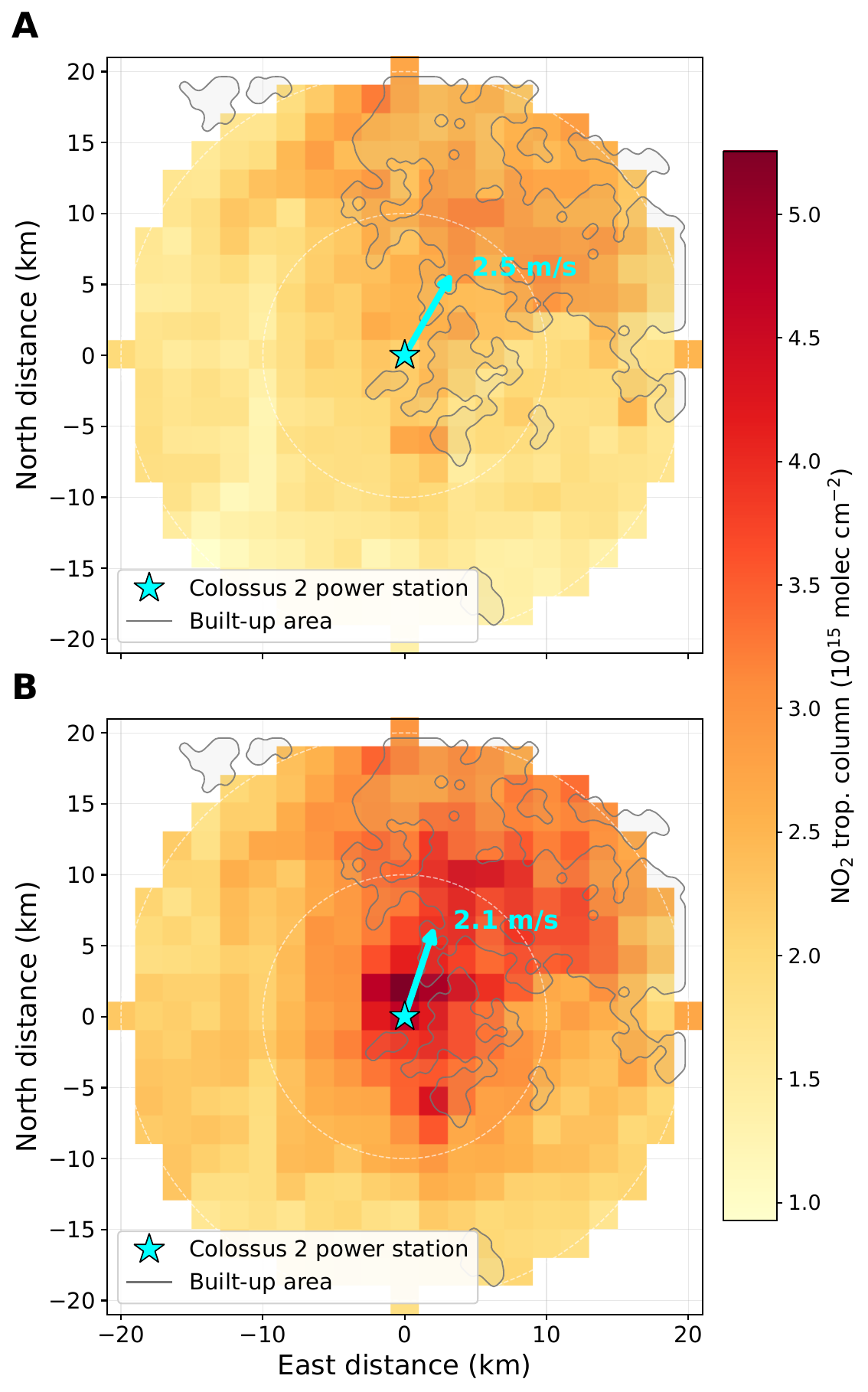}
\caption{Oversampled TEMPO \ce{NO2} columns within 20 km of the Colossus 2 power plant (–90.0398°E, 34.9809°N) for the periods \textbf{A} April–July 2025, before operations began, and \textbf{B} April–July 2026, after operations began. The cyan arrows are period-average, PBL-average wind vectors. Urban/built-up areas are shaded in gray.}
\label{fig:fig1}
\end{figure}

Figure \ref{fig:fig1} shows oversampled TEMPO \ce{NO2} columns observed within 20 km of the Colossus 2 power plant on a 2×2-km$^{2}$ grid, both before and after operations began. Pre-operations, column concentrations were highest to the northeast, with peak values of $3.2\times10^{15}$ molec cm$^{-2}$ over the Memphis urban area. Post-operations, they are highest immediately north of the power plant, in line with the mean wind vector and peaking at $5.3\times10^{15}$ molec cm$^{-2}$, and the mean \ce{NO2} column within a 10 km radius increases by 50\%, from $2.2\times10^{15}$ to $3.3\times10^{15}$ molec cm$^{-2}$. CEMS data for the TVA Southaven Power Plant, situated just $\sim$1.5 km north of the Colossus 2 plant, indicate that it emitted only $\sim$20 kg \ce{NO_x} h$^{-1}$ on average during the two periods, and according to the High Resolution Rapid Refresh (HRRR; \cite{dowell2022}) meteorological analysis product, the regional vector-average planetary boundary layer (PBL) wind speed decreased by only $\sim$15\% between periods, from 2.5 to 2.1 m s$^{-1}$, suggesting that weaker ventilation cannot explain the change. We therefore attribute the stark increase in local \ce{NO2} concentrations observed in Fig. \ref{fig:fig1} to gas turbine operations at the Colossus 2 power plant.

To assess the magnitude of emissions from Colossus 2 over time, we average the hourly TEMPO observations at two-week intervals from August 2025 to August 2026, rotating cloud-free snapshots by HRRR wind directions and discarding scenes with PBL wind below 3 m s$^{-1}$. Wind rotation aligns plumes that may be undetectable on a single pass, preserving plume signal while attenuating background noise \cite{valin2013}. We remove urban background \ce{NO2} to isolate the power plant plume by subtracting the mean two-week-average wind-rotated retrieval for December 2024 to June 2025, before turbines were deployed (Extended Methods). Mean plumes fall along the horizontal axis, and we estimate their source rates using a CSF method that accounts for plume chemistry following \cite{gauld2026}. The method estimates apparent \ce{NO2} source rates from cross-plume line integrals at different distances downwind. It returns the mean of these estimates after discarding values below the 80$^\text{th}$ percentile, removing estimates that are biased low in the near-field due to ozone titration, which delays \ce{NO2} formation \cite{varon2024}, and in the far-field due to \ce{NO_x} oxidation \cite{valin2013}. To convert from \ce{NO2} to a \ce{NO_x} source rate, we use a fixed \ce{NO_x}-to-\ce{NO2} ratio of 1.38$\pm$0.10 \cite{bierle2023}, assuming photochemical steady-state conditions apply for mid-field estimates \cite{gauld2026}.

\begin{figure}%[tbhp]
\centering
\includegraphics[width=\linewidth]{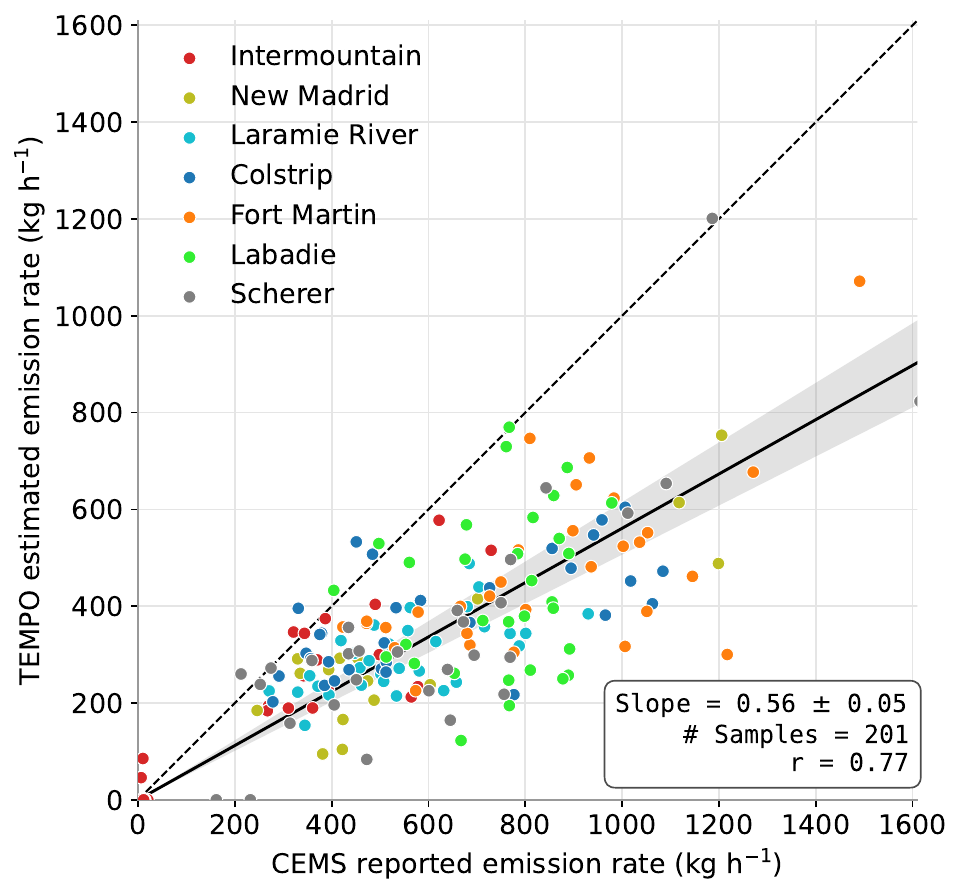}
\caption{Calibration of two-week-average TEMPO source-rate estimates against CEMS data from seven US power plants from December 2024 through June 2026, for two-week periods with mean CEMS rates \textless 1500 kg h$^{-1}$. Gray shading represents uncertainty in the best-fit slope (Extended Methods).}
\label{fig:fig2}
\end{figure}

We calibrate this source-rate retrieval against CEMS data at US power plants. Figure \ref{fig:fig2} compares our TEMPO source-rate estimates with the corresponding two-week-average CEMS values for seven power plants from December 2024 through June 2026. Since TEMPO does not observe a pre-operational period for these power plants, and since they have no significant nearby urban emissions, here we subtract the background as the mean upwind column in the time-averaged scene. Our retrieval underestimates CEMS by a factor of 1.78$\pm$0.14 on average (1.44–1.89 per site), consistent with other reports of low bias in TEMPO \ce{NO_x} source rate retrievals for large power plants \cite{sun2025,beaudry2026}. Potential contributors include errors in the \ce{NO_x}-to-\ce{NO2} ratio, plume masking, and/or the TEMPO \ce{NO2} column retrievals. Since the correlation is high (r = 0.77),  we use the best-fit line to correct our source rate estimates for Colossus 2 (Extended Methods).

\begin{figure*}%[tbhp]
\centering
\makebox[\linewidth][c]{\includegraphics[width=1.3\linewidth]{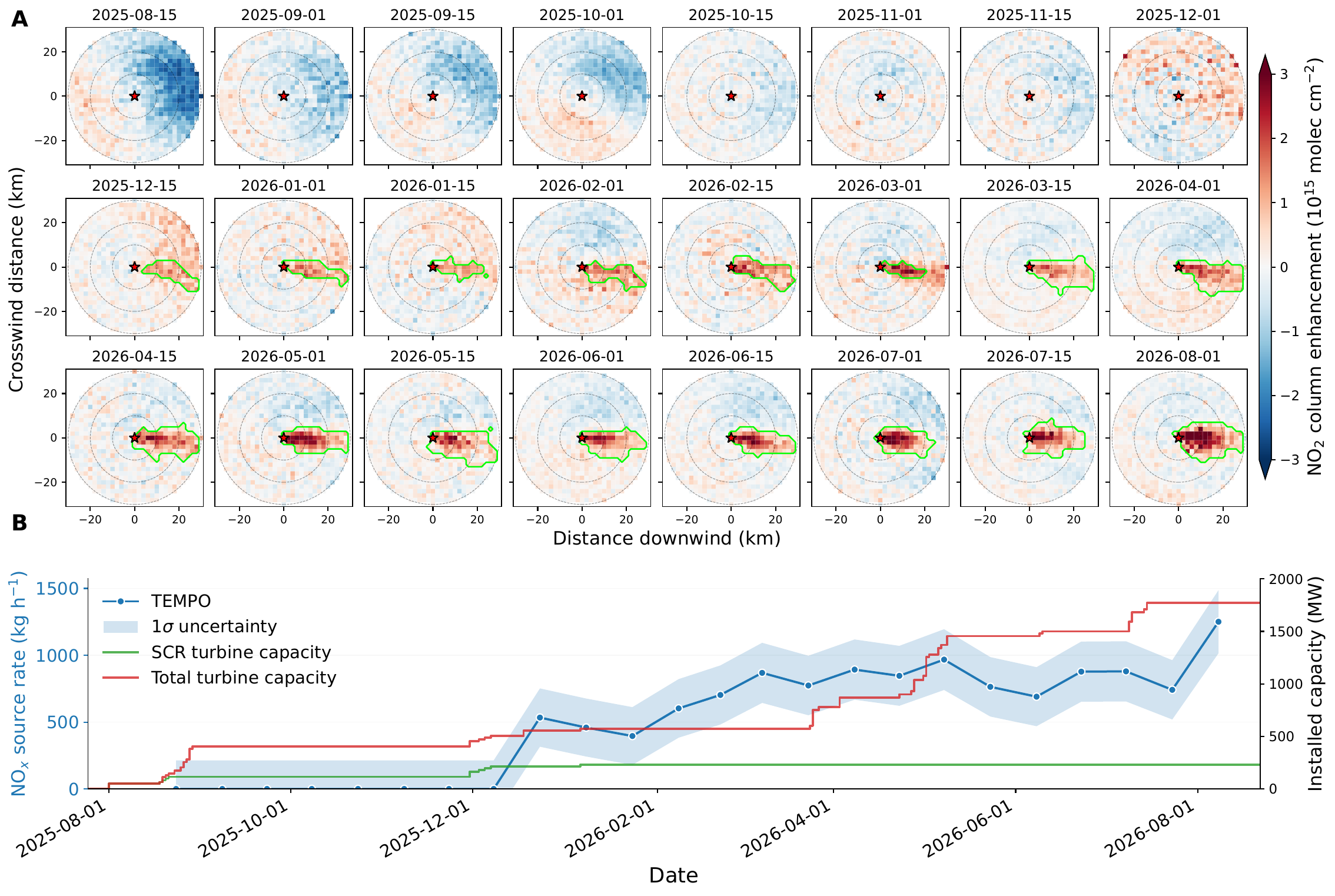}}
\caption{Two-week average NO$_2$ plumes and NO$_x$ source rates from the Colossus 2 power plant from 15 August 2025 to 15 August 2026. \textbf{A} Wind-rotated NO$_2$ enhancements above background, with plumes delineated in green, labeled by period start date. Red stars mark the power plant. \textbf{B} Two-week-average NO$_x$ source rates inferred from \textbf{A} plotted at period mid-point, alongside reported values for the turbines’ total and SCR-equipped nameplate capacity \cite{MCEQ2026}.}
\label{fig:fig3}
\end{figure*}

Figure \ref{fig:fig3} shows the resulting two-week-average wind-rotated \ce{NO2} images and associated \ce{NO_x} source rates from August 2025 through August 2026, along with deployed turbine capacity over time as reported to the Mississippi Commission on Environmental Quality \cite{MCEQ2026}. A plume was first detected in December 2025, at a rate of 530$\pm$220 kg \ce{NO_x} h$^{-1}$ (mean $\pm$ error standard deviation). Emissions appear to lag behind capacity buildout, indicating partial operation of the deployed fleet over time. They were markedly higher in February and later months, peaking at $\sim$1250$\pm$235 kg h$^{-1}$ in August and averaging 870$\pm$165 kg h$^{-1}$ after February 2026. This average is $\sim$19$\pm$4 (95\% CI 11--26) times higher than outlined in the permit for 41 permanent SCR-equipped turbines under best available control technology (BACT) requirements (47 kg h$^{-1}$; \cite{MDEQ2026}; Extended Methods); emissions at that rate would be undetectable by our TEMPO analysis. According to CEMS data, monthly emissions from the nearby TVA Southaven Power Plant averaged only 21$\pm$3 kg h$^{-1}$ during the time period of Fig. \ref{fig:fig3}, peaking at 25 kg h$^{-1}$ in August 2025. Comparing our post-February average source rate with the Colossus 2 turbine deployment history, we infer an emission intensity of 0.93$\pm$0.18 g \ce{NO_x} kWh$^{-1}$ for non-SCR-equipped turbines.

Our study demonstrates that satellites can detect and quantify \ce{NO_x} emissions from AI data centers powered by non-SCR-equipped gas turbines, enabling independent monitoring of air pollution and compliance with environmental regulations. Similar analysis could be conducted using \ce{NO2}-sensitive instruments with global coverage in low-Earth orbit. As AI data centers expand to multi-gigawatt scales, with facilities approaching 5 GW already under development \cite{smith2026} and $\sim$140 GW of new turbine capacity announced in the US alone \cite{EIP2026}, dedicated gas power generation could become an increasingly important source of air pollution. Satellite observations provide a capability to monitor these emerging sources worldwide.

\clearpage

\clearpage

\section*{Extended Methods}

\section*{TEMPO Time Averaging Analysis}

We use the TEMPO L2 V04 tropospheric \ce{NO2} vertical columns for our time averaging analysis, including only observations with the highest per-pixel quality flag (QA value of 0) and effective cloud cover below 0.2. For a baseline estimate, we compute non-rotated time averages from April through July for 2024, 2025, and 2026. In 2026, this is a period of sustained high emissions from the Colossus 2 power plant with available CEMS data from the nearby TVA Southaven Power Plant. The vector-averaged PBL wind speeds for those periods are 1.9, 2.5, and 2.1 m s$^{-1}$, with peak enhancements of $4.0 \times 10^{15}$, $3.2 \times 10^{15}$, and $5.3 \times 10^{15}$ molec cm$^{-2}$, respectively. Between 2024 and 2025, the decrease in peak enhancement by $\sim$20\% is consistent with the $\sim$32\% increase in mean wind speed, and the peak enhancement falls about 8 km north and 6 km east of the Colossus 2 power plant. In 2026, the peak enhancement falls in the pixel immediately north of Colossus 2, and the 66\% increase in peak enhancement cannot be explained by the $\sim$15\% decrease in wind speed. 

For wind-rotated averages, we define two-week intervals for each month from December 2024 through August 2026, between the 1$^\text{st}$ and $15^\text{th}$ and between the 15$^\text{th}$ and 1$^\text{st}$ of the next month. We collect all non-cloudy \ce{NO2} fields over the source location, as well as the local HRRR wind speed and direction at the source. We then filter all granules for 3 m s$^{-1}$ minimum wind speed and rotate the scene to horizontally align wind directions. 

To convert vertical columns to enhancements above background, we must subtract the background from the retrieval field. For our power plant CEMS analysis, which focuses on seven remote point sources, we define the background as the average vertical \ce{NO2} column in pixels upwind of the source after two-week-average wind rotation, following a similar approach to \cite{valin2013}. However, in the urban environment of Colossus 2, other nearby sources can also enhance the downwind signal, so that a uniform background subtraction is unreliable. Therefore, for our Colossus 2 plume analysis, we use wind-rotated TEMPO observations from before the emissions began to define a 2D background field, which we then subtract from post-operations wind-rotated scenes. To account for temporal (e.g., seasonal) variability in the magnitude of the background field, before subtraction, we also scale the background such that the upwind columns have the same average value as those in the target retrieval field.

To create plume masks for the wind-rotated two-week-average retrieval fields, we apply a total variation filter and then mask downwind enhancements that are at least 2–3 standard deviations above the background noise floor. We require masks to begin within 10 km of the source location, discard mask regions smaller than 5 pixels, and remove regions outside the downwind section of the scene ($\pm$25° wedge from source location). We then apply a sequence of erosion and dilation operations to ensure the final mask is relatively smooth and connected. We use the cross-sectional flux (CSF) method described by \cite{gauld2026}, which infers source rates from cross-plume \ce{NO2} line integrals [kg m$^{-1}$] at different distances. We multiply these line integrals by the average HRRR PBL wind speed across all TEMPO observations and a \ce{NO_x}-to-\ce{NO2} ratio of 1.38±0.10 \cite{bierle2023}, representing photochemical steady-state (PSS) conditions, to produce a downwind profile of source-rate estimates. These estimates are expected to be biased low in the near-field of the source, where ozone titration by the emitted \ce{NO} delays \ce{NO2} formation, and in the far-field due to \ce{NO_x} oxidation to \ce{HNO3}. To account for this, we discard estimates below the 80$^\text{th}$ percentile and report the mean of the remaining (mid-field) estimates. Our results are insensitive to the choice of percentile threshold; varying it between the 70$^\text{th}$ and 90$^\text{th}$ percentiles produces a spread of \textless 1\% in the final source-rate estimates for Colossus 2.

\section*{Source Rate Retrieval Error}

Errors in satellite-derived \ce{NO_x} source rates arise from uncertainties in the \ce{NO2} column retrieval, wind speed, background subtraction, \ce{NO_x}-to-\ce{NO2} ratio, and source-rate retrieval method. These errors can be random or systematic. The random component is primarily from wind speed and can exceed 50\% ($1\sigma$) for a single pass \cite{gauld2026}. This error is reduced when averaging multiple estimates over time, and we find total random error \textless 20\% for our post-February-2026 source-rate estimate. Systematic error in our estimates averages 44\% (low bias) relative to CEMS. Other TEMPO studies report similar low biases using a variety of quantification methods. Sun et al. \cite{sun2025} derived source rates for 14 large power plants from TEMPO V03 \ce{NO2} columns using the directional derivative approach and found a larger low bias of 75\% (a factor of 4), which they attributed to systematic errors in the assumed \ce{NO_x}-to-\ce{NO2} ratio and in the V03 retrieval product. More recent work by Beaudry et al. \cite{beaudry2026} shows a smaller $\sim$30\% low bias in TEMPO source-rate estimates for large power plants using the ``rise-decay'' method with a signal-derived retrieval (SDR) product. \ce{NO2} column concentrations in the SDR product are up to 25\% greater than in the TEMPO V04 product near strong sources \cite{beaudry2026}, implying an expected TEMPO-CEMS scale factor of 0.7/1.25 = 0.56 (i.e., up to 44\% low bias) for TEMPO V04, consistent with our result. Beaudry et al. \cite{beaudry2026} also identify \ce{NO_x} lifetime and \ce{NO_x}-to-\ce{NO2} ratio as important sources of error in the rise-decay method, which considers TEMPO observations along the full length of the plume. In contrast, our CSF approach considers only observations in the mid-field of the plume, where PSS conditions would apply, obviating the need to prescribe or fit the \ce{NO_x} lifetime and the dependence of the \ce{NO_x}-to-\ce{NO2} ratio on distance.

We characterize random errors in our two-week-average source-rate retrievals using the results of the CEMS calibration. Error in the best-fit slope $a$ (here for the line $Q_\text{CEMS} = a \times Q_\text{TEMPO}$, the inverse of the line shown in Fig. 2) is computed as the standard deviation of the per-site slope estimates, and random error in the individual estimates is characterized as the standard deviation of fit residuals. These two error terms are summed in quadrature, and the final source rate estimate $Q$ is given by

\begin{equation}
Q = a\hat{Q} \pm \sqrt{\left(\hat{Q}\,\sigma_a\right)^2 + \sigma_{\mathrm{resid}}^2}.
\label{eq:emission_uncertainty}
\end{equation}

where $\hat Q=489$ kg h$^{-1}$ is the initial CSF estimate before correction by the CEMS calibration slope $a=1.78$, $\sigma_a = 0.14$ is the estimated 1$\sigma$ uncertainty in $a$, and $\sigma_\text{resid} = 213 \text{ kg h}^{-1}$ is the standard deviation of the fit residuals. When averaging over multiple two-week periods, we use the same quadrature summation in Eq. \ref{eq:emission_uncertainty}, but use the standard deviation across samples $\sigma_s$ in place of $\sigma_\text{resid}$. For our average post-February 2026, we find $\sigma_s = 150 \text{ kg h}^{-1}$ across 11 samples. When combined with our slope calibration error $\hat Q \sigma_a=68$ kg h$^{-1}$, we obtain a total $1\sigma$ error of 165 kg h$^{-1}$ on the period mean (870$\pm$165 kg h$^{-1}$). Dividing by expected emissions of 47 kg h$^{-1}$ yields a factor of 19$\pm$4 (95\% confidence interval 11–26).

\section*{Expected \ce{NO_x} emissions and reported turbine power output}

Our figure for expected \ce{NO_x} emissions of 47 kg h$^{-1}$ under best available control technology (BACT) requirements is based on public records from the Mississippi Department of Environmental Quality (MDEQ). The MDEQ permit to construct 41 permanent turbines granted on March 11, 2026 \cite{MDEQ2026} lists three turbine blocks (AA-000, AB-000, and AC-000) spanning three different turbine models. All are natural gas-fired simple cycle combustion turbines equipped with Selective Catalytic Reduction (SCR) and oxidation catalyst (oxcat). The BACT emission limits for the three blocks are listed at 1.45, 3.05, and 3.74 lb h$^{-1}$ (0.66, 1.38, 1.70 kg h$^{-1}$) respectively for AA-000, AB-000, and AC-000, for a 3-hour rolling average excluding startup and shutdown, with corresponding annual limits of 5.60, 11.38, and 15.47 short tons a$^{-1}$ (0.58, 1.18, 1.60 kg h$^{-1}$) including startup and shutdown. Summed over all 41 turbines, this yields a fleet total of 46.9 kg h$^{-1}$ (3-hour basis) or 41.6 kg h$^{-1}$ (annualized).

Our figures for deployed turbine capacity (Fig.~3) are based on public records from the Mississippi Commission on Environmental Quality (MCEQ). Information on the Colossus 2 turbines reported to MCEQ includes the unit type, emission controls, nameplate capacity, local expected power output, date added to the site, and retirement date for each unit \cite{MCEQ2026}. In Figure 3, we chart the total nameplate capacity over time based on the date each turbine was added, for all turbines and separately for those with SCR+oxcat systems. The earliest turbine retirement date is August 18, 2026. Turbine capacity is therefore purely additive for our full analysis period from August 15, 2025 through August 15, 2026. Total installed capacity according to MCEQ \cite{MCEQ2026} is 1770.72 MW, with 1540 MW attributed to turbines not equipped with SCR. To compute annualized emissions per kW of capacity post-February 2026, we divide our 870$\pm$165 kg h$^{-1}$ estimate by the time-averaged capacity of installed turbines without SCR over the averaging period (933.5 MW).

{
\footnotesize
\bibliographystyle{unsrt}
\bibliography{references}

\begin{thebibliography}{10}

\bibitem{NAACPvxAI2026}
{NAACP}.
\newblock {Complaint, {NAACP v. X.AI Corp.}}
\newblock No. 3:26-cv-74-MPM-JMV, U.S. District Court for the Northern District
  of Mississippi, 2026.

\bibitem{trinetti2026}
Corey Trinetti.
\newblock {xAI's Colossus Cluster: A Gigawatt of AI Data Centers, Built Off the
  Grid}.
\newblock {\em Measured AI}, 2026.

\bibitem{zoogman2017}
P.~Zoogman, X.~Liu, R.M. Suleiman, W.F. Pennington, D.E. Flittner, J.A.
  Al-Saadi, B.B. Hilton, D.K. Nicks, M.J. Newchurch, J.L. Carr, S.J. Janz, M.R.
  Andraschko, A.~Arola, B.D. Baker, B.P. Canova, C.~{Chan Miller}, R.C. Cohen,
  J.E. Davis, M.E. Dussault, D.P. Edwards, J.~Fishman, A.~Ghulam, G.~{González
  Abad}, M.~Grutter, J.R. Herman, J.~Houck, D.J. Jacob, J.~Joiner, B.J.
  Kerridge, J.~Kim, N.A. Krotkov, L.~Lamsal, C.~Li, A.~Lindfors, R.V. Martin,
  C.T. McElroy, C.~McLinden, V.~Natraj, D.O. Neil, C.R. Nowlan, E.J.
  O'Sullivan, P.I. Palmer, R.B. Pierce, M.R. Pippin, A.~Saiz-Lopez, R.J.D.
  Spurr, J.J. Szykman, O.~Torres, J.P. Veefkind, B.~Veihelmann, H.~Wang,
  J.~Wang, and K.~Chance.
\newblock {Tropospheric emissions: Monitoring of pollution (TEMPO)}.
\newblock {\em Journal of Quantitative Spectroscopy and Radiative Transfer},
  186:17--39, 2017.

\bibitem{MCEQ2026}
{Mississippi Commission on Environmental Quality}.
\newblock {Agreed Order No. 7739-26, In re: {MZX Tech LLC}}.
\newblock Appendix A, 2026.

\bibitem{MDEQ2026}
{Mississippi Department of Environmental Quality}.
\newblock {Permit to Construct Air Emissions Equipment and PSD Authority}.
\newblock Permit No. 0680-00119, issued to MZX Tech LLC, 2026.

\bibitem{CAMPD}
{U.S. Environmental Protection Agency}.
\newblock {Clean Air Markets Program Data (CAMPD)}.
\newblock \url{https://campd.epa.gov/}, n.d.

\bibitem{dowell2022}
David~C. Dowell, Curtis~R. Alexander, Eric~P. James, Stephen~S. Weygandt,
  Stanley~G. Benjamin, Geoffrey~S. Manikin, Benjamin~T. Blake, John~M. Brown,
  Joseph~B. Olson, Ming Hu, Tatiana~G. Smirnova, Terra Ladwig, Jaymes~S.
  Kenyon, Ravan Ahmadov, David~D. Turner, Jeffrey~D. Duda, and Trevor~I.
  Alcott.
\newblock {"The High-Resolution Rapid Refresh (HRRR): An Hourly Updating
  Convection-Allowing Forecast Model. Part I: Motivation and System
  Description"}.
\newblock {\em Weather and Forecasting}, 37(8):1371 -- 1395, 2022.

\bibitem{valin2013}
L.~C. Valin, A.~R. Russell, and R.~C. Cohen.
\newblock {Variations of OH radical in an urban plume inferred from NO2 column
  measurements}.
\newblock {\em Geophysical Research Letters}, 40(9):1856--1860, 2013.

\bibitem{gauld2026}
Kevin~D. Gauld, Daniel~J. Varon, Andrew~K. Thorpe, and Robert~O. Green.
\newblock {Quantifying Nitrogen Oxide Point Sources With the EMIT Satellite
  Imaging Spectrometer}.
\newblock {\em Geophysical Research Letters}, 53(22):e2026GL124107, 2026.

\bibitem{varon2024}
Daniel~J. Varon, Dylan Jervis, Sudhanshu Pandey, Sebastian~L. Gallardo,
  Nicholas Balasus, Laura~Hyesung Yang, and Daniel~J. Jacob.
\newblock {Quantifying NOx point sources with Landsat and Sentinel-2 satellite
  observations of NO2 plumes}.
\newblock {\em Proceedings of the National Academy of Sciences},
  121(27):e2317077121, 2024.

\bibitem{bierle2023}
S.~Beirle, C.~Borger, A.~Jost, and T.~Wagner.
\newblock {Improved catalog of NO$_x$ point source emissions (version 2)}.
\newblock {\em Earth System Science Data}, 15(7):3051--3073, 2023.

\bibitem{sun2025}
Kang Sun, Jobaer~Ahmed Saju, Caroline~R. Nowlan, Gonzalo González~Abad, and
  Xiong Liu.
\newblock {Hourly Nitrogen Oxides Emissions Estimated From TEMPO and Comparison
  With Facility-Level Monitoring Data}.
\newblock {\em Journal of Geophysical Research: Atmospheres},
  130(23):e2025JD044565, 2025.

\bibitem{beaudry2026}
Samuel~J. Beaudry, Qindan Zhu, and Ronald~C. Cohen.
\newblock {Quantification of Power Plant {NO$_x$} Emissions: A Reference for
  Diurnal and Seasonal Accuracy of {TEMPO}}.
\newblock EarthArXiv, 2026.

\bibitem{smith2026}
Matthew~S. Smith.
\newblock {What Will It Take to Build the World's Largest Data Center?}
\newblock {\em IEEE Spectrum}, 2026.

\bibitem{EIP2026}
{Environmental Integrity Project}.
\newblock The power behind {AI}: Wave of dirty gas power plants planned for
  data centers.
\newblock Report, Environmental Integrity Project, Washington, DC, July 2026.

\end{thebibliography}
}

\end{document}